# Mid-Infrared Photothermal Relaxation Intensity Diffraction Tomography for Video-rate Volumetric Chemical Imaging

**DANCHEN JIA,**[1,†] **DASHAN DONG,**[1,†] **TONGYU LI,**[1] **HAONAN ZONG,**[1] **JIABEI ZHU,**[1] **XINYAN TENG,**[2] **LEI TIAN,**[1,2,*] **AND JI-XIN CHENG**[1,2,*]

[1]*Department of Electrical and Computer Engineering, Boston University, Boston, MA, 02215, USA*
[2]*Department of Biomedical Engineering, Boston University, Boston, MA, 02215, USA*
[†]*These authors contributed equally.*
*\*leitian@bu.edu, jxcheng@bu.edu*

**Abstract:** Three-dimensional molecular imaging of living cells is essential for unraveling cellular metabolism and response to therapies. However, existing volumetric methods, including fluorescence microscopy and quantitative phase imaging, either require fluorescent labels or lack chemical specificity. Mid-infrared (mid-IR) photothermal microscopy provides label-free spectroscopic contrast with sub-micrometer resolution but is limited by slow acquisition rates, precluding 3D live-cell studies. Here, we present a photothermal relaxation intensity diffraction tomography (PRIDT) system that encodes mid-IR absorption induced refractive index change via a photothermal relaxation scheme and recovers it through intensity diffraction tomography. PRIDT achieves video-rate volumetric chemical imaging with up to 15 Hz per wavelength and offers lateral and axial resolutions of 264 nm and 1.12 µm over a volumetric field of view of 50×50×10 µm³. We showcase high-speed PRIDT imaging of protein and lipid metabolism in ovarian cancer cells and lipid-droplet dynamics in live cells. PRIDT opens new avenues for rapid, quantitative, three-dimensional molecular imaging in living systems.

## 1. Introduction

Three-dimensional (3D) visualization of biological structures and functions at the subcellular level is essential for better understanding of biological processes such as cancer metabolism, drug response, and disease progression. Bright-field and phase-contrast microscopy offers basic morphological insights but lacks molecular information. Fluorescence microscopy addresses this limitation and has been developed with a range of 3D modalities, including confocal [1,2], two-photon [3,4], light-sheet [5–7], light-field [8,9]. Despite their high specificity and resolution, fluorescence microscopy require labeling, which complicates sample preparation, limits throughput, and may perturb native molecular functions.

Stimulated Raman scattering (SRS) and coherent anti-Stokes Raman scattering (CARS) have been widely used for label-free chemical imaging with bond specificity [10–13]. To further accelerate 3D SRS imaging, Bessel beam-based stimulated Raman projection tomography [14] and remote-focusing SRS microscope [15] have been introduced, enabling quantification of the total chemical composition of 3D biological specimens. More recently, the integration of SRS microscopy with expansion microscopy has enabled nanoscale 3D chemical imaging [16]. Despite these advances, the imaging speed of coherent Raman techniques remains fundamentally constrained by the intrinsically weak spontaneous Raman scattering cross-section of $10^{-30}$ cm$^2$ for most biomolecules [17].

Infrared (IR) microscopy offers higher signal levels than Raman scattering due to its intrinsically stronger linear absorption cross-section of $10^{-18}$ cm$^2$. Mid-infrared photothermal (MIP) microscopy employs a visible probe beam to detect IR absorption–induced refractive index changes, thus achieving 3D chemical imaging with sub-micrometer spatial resolutions in a scanning configuration [18–22], and enabling functional analysis in life science and material science [23–27]. Wide-field MIP microscopy improves MIP imaging speed using a virtual lock-in camera approach [28–31]. Wide-field infrared-excited fluorescence-detected microscopy further enables cellular imaging beyond video rate [32]. Furthermore, double-shot volumetric chemical imaging has been demonstrated by integrating fluorescence-detected MIP [33, 34] with Fourier light field microscopy [35], while the chemical contrast here still relies on fluorescent labeling.

In parallel to chemical microscopy, quantitative phase imaging (QPI) has been developed to measure subtle phase shifts in transparent biological samples without labeling [36]. Optical diffraction tomography (ODT) extends QPI to three dimensions using multi-angle illumination [37–40]. However, the laser coherence required for ODT introduces strong speckle noise. Intensity diffraction tomography (IDT) has been demonstrated, enabling 3D refractive index reconstruction from intensity-only measurements with partially coherent illumination without speckle articles [41–43]. QPI enables noninvasive, high-resolution, long-term imaging of organelle dynamics, but its contrast arises from refractive-index variations lacking molecular specificity. To address this limitation, QPI and MIP microscopy are integrated to encode the chemical information of subcellular organelles into phase contrasts [44]. To further explore label-free volumetric chemical imaging, the MIP effect is incorporated into ODT [45, 46] and IDT [47] to map chemicals in cells and *C. elegans*. Thus far, these methods have only been demonstrated on fixed samples in deuterium oxide, with a 3D chemical imaging speed limited to approximately 20 seconds per volume. Real-time imaging of living cells remains a challenge.

We recognize that volumetric MIP imaging in a wide-field configuration is fundamentally limited by the slow thermal relaxation dynamics of aqueous environments [48–50]. To overcome this limitation, we introduce photothermal relaxation intensity diffraction tomography (PRIDT) that decouples photothermal relaxation time from image acquisition speed through transient-state sampling. Specifically, this approach leverages the slow thermal relaxation of bulk water by introducing two fine-tuned pump–probe delay times to capture transient phase changes associated with the IR absorption of subcellular structures. Moreover, by precisely modulating laser pulse timing and fluence, PRIDT maintains thermal equilibrium near physiological temperatures, ensuring compatibility with live-cell imaging. We applied PRIDT to perform 3D mid-infrared hyperspectral imaging of cancer cells in the fingerprint region to investigate cancer metabolism. Azide-tagged fatty acids were used to visualize fatty acid uptake at the subcellular level. Finally, video-rate PRIDT imaging of lipid droplet dynamics in live cancer cells was achieved at 15 Hz per volume.

## 2. Results

### 2.1 PRIDT system

The PRIDT system is a dual-delay, pump-probe, multi-angle imaging system capable of volumetric chemical imaging at video-rate speed. Optical layout is shown in Fig. 1(a). Nanosecond mid-IR pulses (900–2300 cm$^{-1}$) serve as the pump beam. A custom-built ring-shaped array of 450-nm fiber-collimated laser diodes modulated into nanosecond pulses provides wide-field oblique illumination at $N_{\text{illum}}$ = 16 angles. The transmitted probe light, modulated by the sample's photothermal response, is collected by a 60× water-immersion objective with an adjustable aperture to match the illumination numerical aperture (NA) of 0.9 (see Methods). To sustain high signal-to-noise ratio (SNR) at the short exposures required for the high-speed imaging, we use a camera with 2-million-electron full-well capacity. A kHz-frame-rate camera ($f_{\text{camera}}$ ≤ 1 kHz) operation, yielding a volumetric imaging speed of $f_{\text{camera}}/2N_{\text{illum}}$.

Considering that the thermal relaxation time of bulk water is on the millisecond time-scale, a slow laser repetition rate (< 1 kHz) would favor virtual lock-in detection of purely transient photothermal signals (see Supplement 1, Fig. S1a) [46]. In our practice, however, filling the camera's full-well without excessive per-pulse energy necessitates a high probe repetition rate; we therefore operated probe at 50 kHz with 500-ns pulses, which also mitigates pulse-to-pulse fluctuations. Under these conditions, cumulative water heating dominates the bond-specific transients (see Supplement 1, Fig. S1b). To recover chemically selective contrast, we implement a photothermal relaxation scheme [Fig. 1(b)]. After previous millisecond-scale heating the sample reaches a quasi-steady baseline temperature $T_0$, further absorption from the mid-IR pump pulses induce same transient local temperature rises of the molecular absorbers and water, followed by heat dissipation into the surrounding medium. This relaxation leads to distinct temporal signatures between absorbers (red) and the background medium (gray), due to the differences in the thermal conductivity and heat capacity of the materials. Thus, mid-IR absorption of the target absorbers ($I_{abs}$) can be extracted by subtraction of the photothermal signals at two temporal states ($I_{t1}$, red window; $I_{t2}$, gray window), where the temperature of the background medium remains the same.

$$I_{abs} \propto \Delta T_{abs} - \Delta T_{medium} = S_1 - S_2 = I_{t_1} - I_{t_2} \qquad (1)$$

Here, $\Delta T_{abs}$ is the temperature rise from mid-IR absorption by the absorbers, and $\Delta T_{medium}$ is the temperature rise of the surrounding medium induced due to mid-IR absorption of water and thermal diffusion in the surrounding medium.

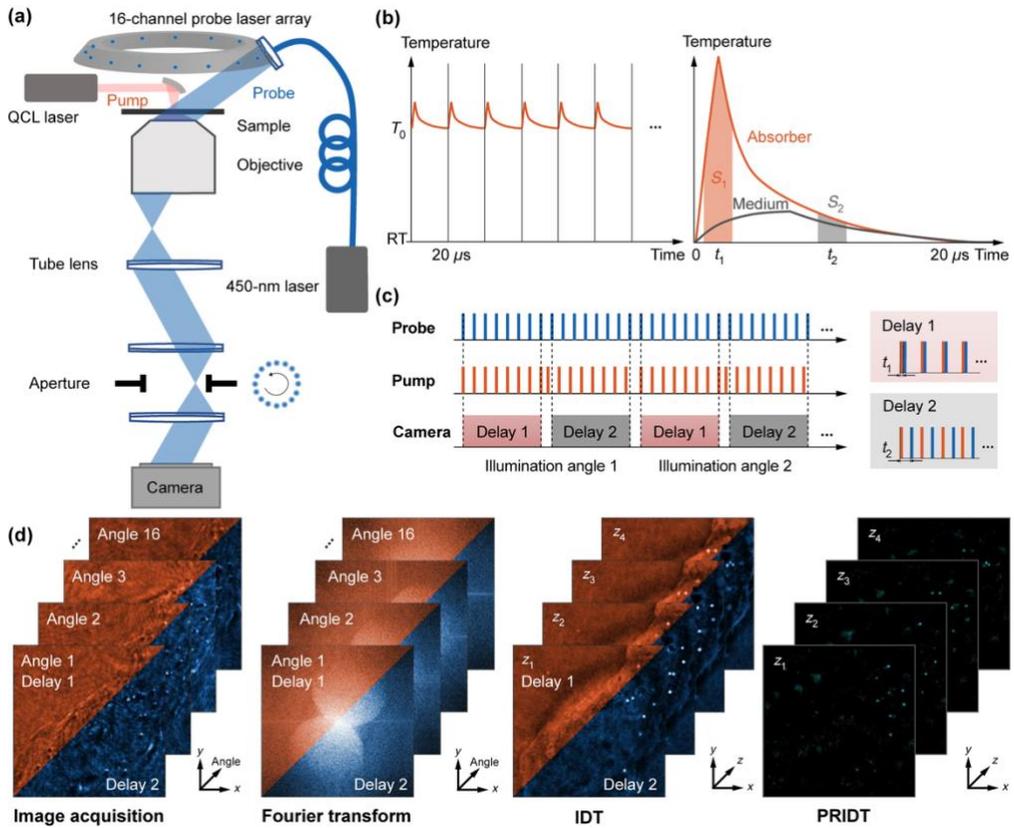

**Fig. 1.** PRIDT system design, timing diagram and data pipeline. (a) Schematic of the PRIDT setup, consisting of a nanosecond mid-IR QCL pump laser, a multi-angle laser diode ring array for tomographic visible probe illumination, and a kHz-frame-rate camera. (b) Thermal dynamics of PRIDT: left, thermo-equilibrium at an elevated environmental temperature $T_0$; right, absorber (red) and medium (gray) responses within a single pump–probe cycle. (c) Timing diagram of mid-IR pump, visible probe and camera exposure synchronized to capture two thermal relaxation states at delays $t_1$ and $t_2$. (d) Image acquisition and reconstruction pipeline: 16-angle illumination images recorded at two delays, IDT volumes reconstructed at each state, and PRIDT volume retrieved by subtracting the two IDT reconstructions.

To achieve video-rate volumetric imaging with this scheme, we use time-multiplexed acquisition [Fig. 1(c)]. The 16 probe angles illuminate sequentially while the pump alternates between chemical resonant states ($t_1$) and reference states ($t_2$). One volume is acquired in 32 frames, 16 for each state, enabling 15 volumes per second (see details in Methods). The data pipeline of PRIDT is outlined in Fig. 1(d). Raw probe images are collected from 16 incident angles and two delay states. Each 16-angle IDT dataset is used to perform tomographic reconstruction of refractive index (RI) distributions via the transfer function-based IDT reconstruction[41]. Incorporating photothermal relaxation dynamics at each angle further provides chemically selective volumetric reconstructions with PRIDT.

## 2.2 Thermal dynamics and spectroscopic imaging of polymer beads

To demonstrate the volumetric chemical imaging capability with PRIDT, we first characterized the transient photothermal response of polymer microspheres embedded in water. Fig. 2(a) shows the temporal evolution of the refractive index change ($\Delta n$) following mid-IR excitation, measured by scanning the dual pump–probe delay windows between resonant states and reference states $t_1$ - $t_2$ [illustrated in Fig. 1(b) and 1(c)]. To accommodate the camera frame rate (1 kHz) with nanosecond-scale photothermal relaxation, a multiplexed pulse generator was utilized to trigger the mid-IR laser and camera from the probe laser array driver's 50 kHz master clock. PMMA microspheres with a diameter of 2 μm exhibited a rapid thermal rise and subsequent relaxation with a time constant of 2.5 μs, whereas the water background showed a slower thermal response, demonstrating the capability of suppressing water background with the dual-delay method. The extracted spectral response [Fig. 2(b)] revealed a strong absorption peak at 1728 cm$^{-1}$, consistent with the C=O stretching mode of PMMA, confirming chemical specificity of the PRIDT signal.

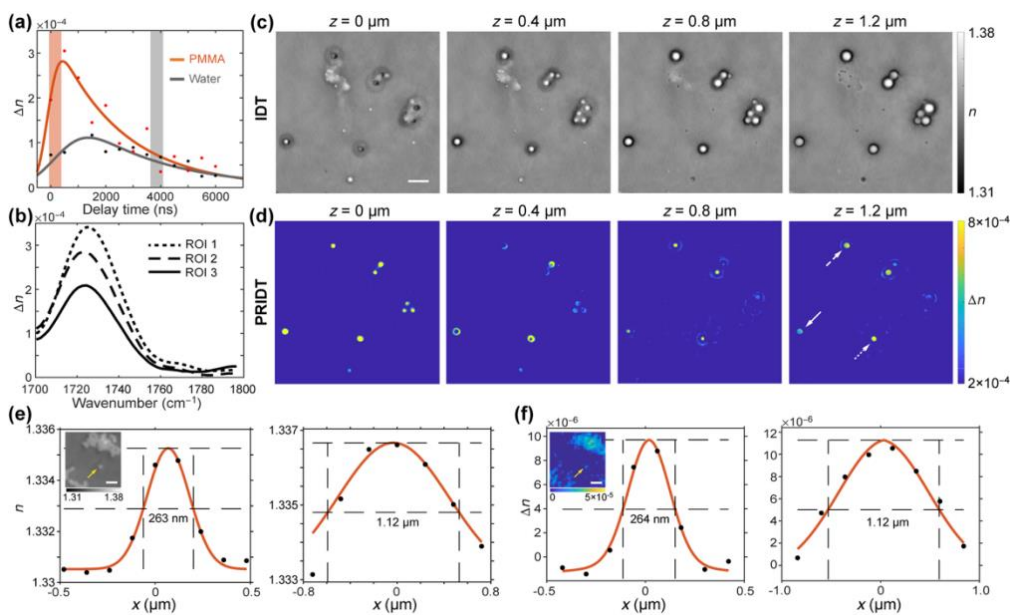

**Fig. 2.** Thermal dynamics, spectral response and image reconstruction of PMMA beads in water. (a) Thermal dynamics of 2 μm-diameter PMMA beads in water (red, PMMA; gray, water; curves fitted by the convolution of the Gaussian instrumental response function and the exponential decay function of the sample). (b) Spectra of PMMA beads (peak at 1728 cm$^{-1}$) indicated by white arrows in the PRIDT images. (c, d) Depth-resolved IDT (c) and PRIDT (d) reconstructions of μm-diameter PMMA beads in water. Scale bar, 5 μm. (e, f) Line profiles of IDT (e) and PRIDT (f) of 200 nm-diameter PMMA beads (highlighted by yellow arrows), with the full width at half maximum (FWHM) marked. Scale bar, 1 μm.

We further evaluated 3D chemical imaging performance in Fig. 2(c) and 2(d). IDT images at multiple axial planes (z = 0 to 1.2 μm) captured the structural distribution of the PMMA microspheres but lacked chemical selectivity. PRIDT reconstructions at corresponding depths revealed clear localization of PMMA beads with strong photothermal contrast, enabling chemical distinction from the surrounding medium [Fig. 2(d)]. To assess resolution and sensitivity, we imaged PMMA nanospheres with a diameter of 200 nm. We further quantified the spatial resolution of IDT [Fig. 2(e)] and PRIDT [Fig. 2(f)]. Line profiles across isolated particles demonstrated lateral and axial full-width half-maximums (FWHM) of 264 nm and 1.12 µm, respectively, governed by optical diffraction limit and photothermal diffusion length (see details in Supplementary Information). These results establish PRIDT as a label-free platform for volumetric, chemically specific imaging with sub-micrometer spatial resolution.

### 2.3 Fingerprint-region spectroscopic imaging of cancer cells

We applied PRIDT to map the chemically resolved subcellular composition of fixed OVCAR-5 ovarian cancer cells. 3D IDT reconstructions across multiple depths [Fig. 3(a)] revealed the subcellular morphology but lacked molecular specificity. By contrast, PRIDT reconstructions at protein and lipid vibrational bands [Fig. 3(b) and 3(c)] provided chemically selective volumetric maps. Hyperspectral image data processing was performed by a least-squares fitting algorithm detailed in Methods section. Protein signals were observed throughout the cytoplasmic region, whereas lipid contrast were localized to lipid droplets.

To further validate spectral fidelity, we extracted hyperspectral PRIDT responses from different regions of interest (ROIs). Spectra obtained from protein-rich regions [Fig. 3(d)] exhibited two peaks around 1552 cm$^{-1}$ and 1656 cm$^{-1}$, corresponding to the amide II and amide I vibrational band, characteristic of cellular protein with $\alpha$-helix secondary structures. In contrast, spectra from lipid-rich regions [Fig. 3(e)] showed a pronounced absorption peak at 1744 cm$^{-1}$, consistent with the C=O stretching mode of lipid esters. These spectral features match established biochemical fingerprints, confirming the molecular contrast provided by PRIDT. Together, these results demonstrate that PRIDT enables volumetric label-free chemical imaging of cancer cells with subcellular spatial resolution, highlighting its potential for biochemical mapping of cellular architecture in 3D and to study protein and lipid metabolisms in a label-free manner.

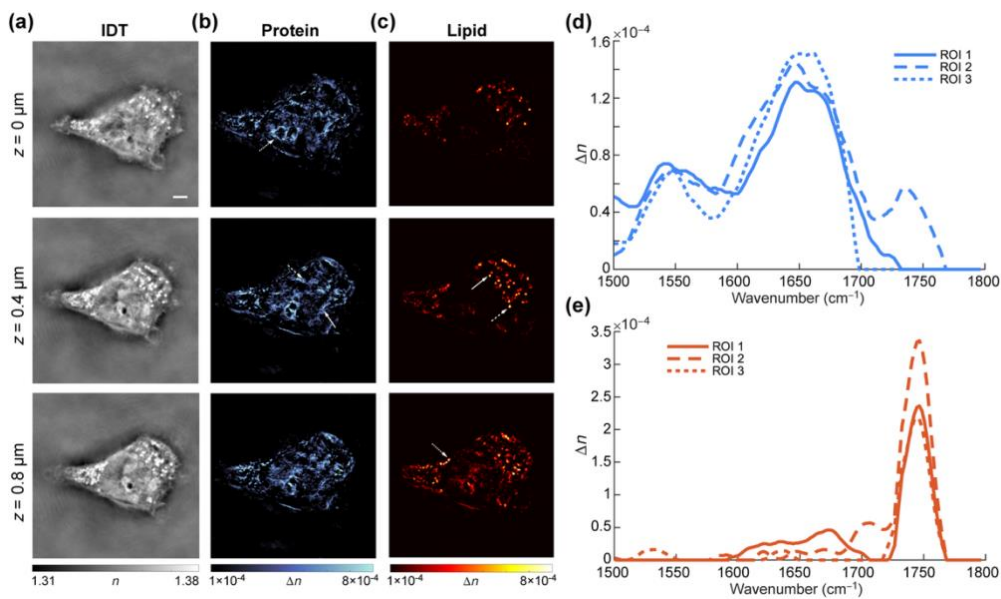

**Fig. 3.** Hyperspectral imaging of fixed cancer cells. (a–c) IDT and PRIDT reconstructions showing 3D chemical maps of proteins (blue) and lipids (hot) in fixed OVCAR-5 cells. Scale bar, 5 µm. (d, e) Representative spectra from ROIs enriched in protein (peaks at 1656 cm$^{-1}$ and 1552 cm$^{-1}$, corresponding to Amide I and Amide II) and lipid ester (peak at 1744 cm$^{-1}$).

*2.4 Imaging fatty acid uptake with PRIDT*

So far, we have shown that PRIDT is capable of distinguishing intrinsic molecules inside cells. Here, we further investigated lipid synthesis in cancer cells using a biorthogonal chemical reporter introduced through metabolic incorporation of azide-tagged fatty acids [51, 52]. These bio-orthogonal tags are widely used to probe dynamic metabolic processes without perturbing native cellular functions. Bai et al. demonstrated MIP imaging of azide-tagged fatty acids in primary neurons [53]. Here, PRIDT reconstructions [Fig. 4(a) and 4(b)] revealed cellular morphology and 3D distribution of lipids at C=O vibrational band, which concentrated within lipid droplets. Concurrently, PRIDT also resolved signals at the azide vibrational band [Fig. 4(c)] of 2096 cm$^{-1}$, with spectral fidelity confirmed in Fig. 4(d) and 4(e). These signals co-localized with lipid droplet regions, confirming the intake of exogenous fatty acids into intracellular lipid pools. The azide contrast further demonstrates the ability of PRIDT to detect non-native vibrational signatures with high sensitivity.

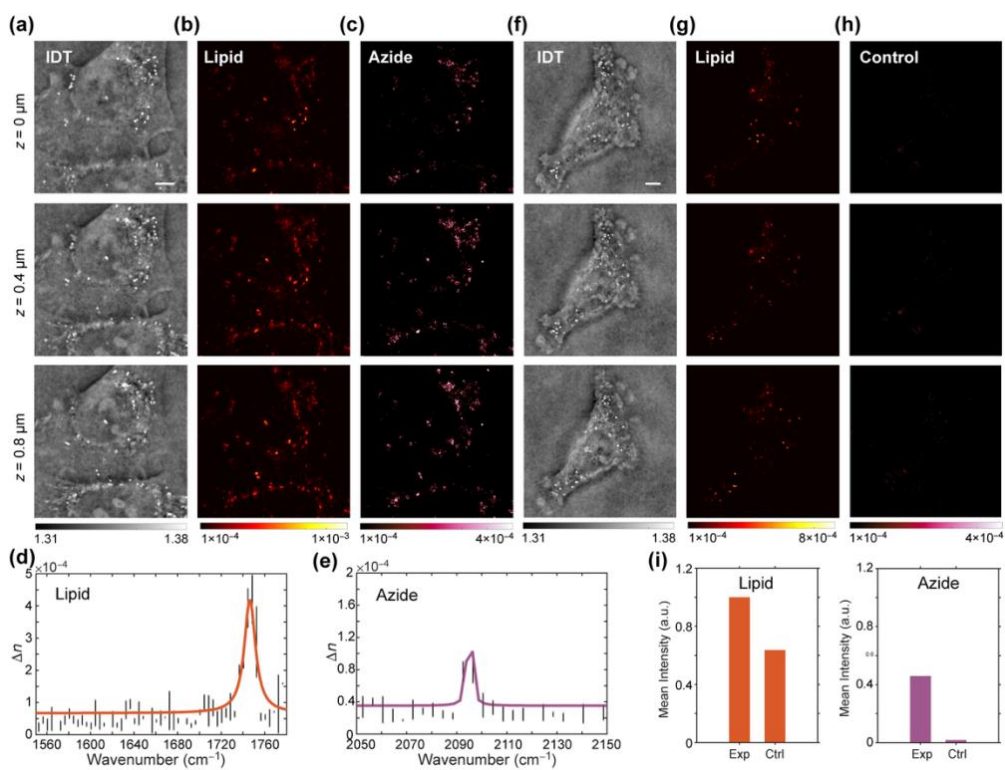

**Fig. 4.** Fatty acid uptake in fixed cancer cells treated with azide-tagged fatty acids. (a–c) IDT (a) and PRIDT images at the lipid (b) and azide (c) channels, shown at different depths of OVCAR-5 cells treated with azide-tagged fatty acids. (d, e) Representative spectra of lipid (peak at 1744 cm$^{-1}$) and azide (peak at 2096 cm$^{-1}$) signals. (f–h) IDT (f) and PRIDT images at the lipid (g) and azide (h) channels for cells without azide treatment (control group). (i) Statistical analysis of the mean lipid (left) and azide (right) signal intensities comparing the experimental and control groups. Scale bar, 5 μm.

As a control, cells without azido-fatty acid treatment exhibited lipid droplet signals but no detectable azide signal [Fig. 4(f–h)], confirming the capability of chemical tracking of PRIDT. Statistical analysis of fatty acid uptake was shown in Fig. 4(i), where the mean intensities of lipid and azide signals were compared to quantify the lipid synthesis from exogenous treatment and endogenous pathways. These results demonstrate that PRIDT enables volumetric monitoring of lipid metabolism in cancer cells, providing an approach for tracing diverse biochemical pathways with three-dimensional subcellular spatial resolution. Moreover, many pharmacological agents contain azide functional groups, such as antiviral azidothymidine as an HIV reverse transcriptase inhibitor [54, 55]. Thus, PRIDT could be further extended to track drug uptake and intracellular distribution at the subcellular level, offering insights for drug discovery and development.

*2.5 Temperature rise in PRIDT*

In PRIDT imaging of living cells, elevated temperatures by water heating raise concerns about potential cytotoxicity. To assess this, we first conducted numerical simulations of heat accumulation in bulky water using a simplified 3D finite element model in COMSOL (see Supplementary Information). The model included a 500-nm PMMA particle located at the $CaF_2$–water interface. As shown in Fig. 5(a), increasing the IR laser repetition rate from 20 kHz to 50 kHz resulted in a rise in equilibrium water temperature from 25 °C to 32 °C.

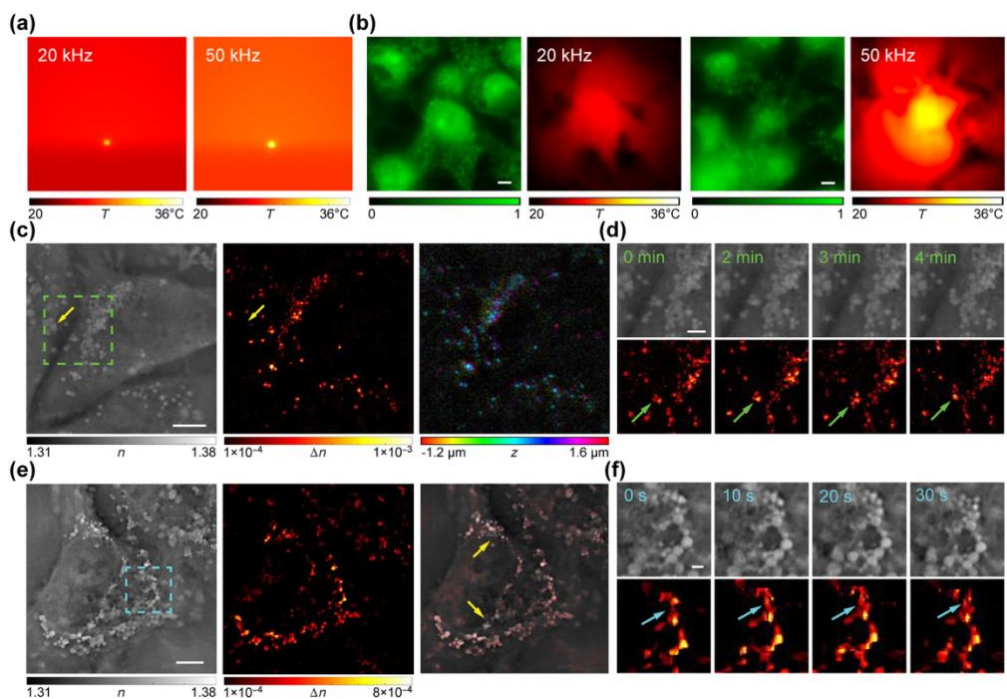

**Fig. 5.** Temperature calibration and PRIDT video of living cells. (a) Simulated temperature maps of photothermal heating of a 500nm-diameter PMMA bead in water at 20kHz and 50kHz. (b) Experimental temperature measurements of cancer cells under photothermal heating at 20kHz (left) and 50kHz (right) using thermo-sensitive fluorescence. Green: fluorescence maps; red: corresponding temperature maps. (c) Representative slices from IDT (left), PRIDT (middle), and the PRIDT z-stack color map of living OVCAR-5 cells. (d) Zoomed-in views of ROI from (c), highlighting lipid droplet dynamics on minute-scale. (e) Representative slices from IDT (left), PRIDT (middle), and their overlay of living OVCAR-5 cells. (f) Zoomed-in views of ROI from (e), showing lipid droplets dynamics on second-scale. Scale bar in (b, c, e), 5 μm; scale bar in (d, f), 1 μm.

To experimentally validate the temperature changes in aqueous cellular environment during PRIDT imaging, we employed FITC-based fluorescence thermometry under the same mid-IR heating conditions (see Methods) [56]. The temperature sensitivity of FITC fluorescence was calibrated to be 1.8% per °C using a heating plate and a precision thermometer (see Supplement 1, Fig. S2). The cellular temperature was then inferred from the decrease in fluorescence intensity [Fig. 5(b)]. The temperature distribution followed the spatial profile of the mid-IR heating beam, with measured temperatures in bulk water reaching ~26 °C under 20 kHz heating pulses and ~33 °C under 50 kHz, at the water absorption peak (1640 $cm^{-1}$). We further extended the measurements across the fingerprint spectral region (see Supplement 1, Fig. S2). These in situ thermometry results confirm that PRIDT imaging elevates the medium temperature to a range close to physiological conditions, which is essential for preserving cellular function and viability.

*2.6 3D Imaging of Lipid Dynamics in Live Cells*

To demonstrate video-rate volumetric chemical imaging in living systems, we applied PRIDT to monitor lipid droplet dynamics in cultured cancer cells at a speed of 15 volumes per second. BM4D denoising was applied to the reconstructed volumes, substantially improving the SNR while preserving structural features. Lipid droplets were clearly identified in the denoised IDT and PRIDT videos [Fig. 5(c), Visualizations 1 to 4]. In Fig. 5(d), we highlighted subcellular areas with evident droplet activity. Notably, in the overlaid image [Fig. 5(e)], several high-refractive-index features did not correspond to lipid signals, likely attributed to lysosomes or other organelles highlighted with yellow arrows. This observation emphasized the chemical specificity of PRIDT over IDT, enabling differentiation between lipid droplets and other refractive subcellular structures.

Time-lapse imaging over 30 s revealed continuous droplet growth and trafficking. In the region highlighted in cyan, newly formed lipid droplets emerged and gradually increased in contrast, while adjacent droplets migrated along cytoplasmic tracks [Fig. 5(f), top]. Corresponding photothermal signals confirmed the chemical specificity of lipid components [Fig. 5(f), bottom, cyan arrows]. Additionally, shown in Supplement 1, Fig. S3 (green arrows), droplet clusters displayed fusion events and dynamic redistribution within the cytoplasm. These observations illustrate the ability of PRIDT to resolve lipid droplet dynamics in 3D, with sufficient temporal resolution to capture subcellular trafficking in real time.

## 3. Conclusion and Discussion

PRIDT pushed the volumetric imaging speed of label-free 3D vibrational microscopy to a video rate, achieving 15 Hz per volume, with a lateral resolution of 264 nm and an axial resolution of 1.12 μm, and across a 50×50×10 μm$^3$ field of view defined by the mid-IR illumination profile. PRIDT add chemical selectivity to quantitative phase imaging, enabling chemically specific volumetric reconstruction. The photothermal relaxation detection method not only enhanced the 3D chemical imaging speed by 300× compared with previously reported BS-IDT [47], but also preserved physiologically compatible temperatures within the sample environment, which is beneficial for live-cell studies. With these advances, we demonstrated label-free monitoring of cancer cell metabolism, including protein secondary structure and lipid droplet dynamics. Furthermore, PRIDT enabled visualization of fatty acid uptake using azide-tagged fatty acids, extending its ability of specific metabolic studies and drug screening in living cells.

Mechanistically, PRIDT leverages differences in photothermal relaxation between targets and their medium to suppress water background and isolate chemical contrast from target absorbers. Importantly, the photothermal relaxation readout enables high-repetition-rate (>1 kHz) detection with sub-microsecond probe pulses, decoupling signal formation from single pulse energy and thus permitting the use of simpler, lower-cost laser diodes rather than high-energy pulsed sources. Operating at high repetition rates also averages laser intensity fluctuations to reduce laser noise, which improves sensitivity at video rates.

Several limitations and opportunities remain. Firstly, PRIDT relies on the distinct photothermal relaxation characteristics between target structures and their surrounding medium. Therefore, contrast is strongest for dense or high-absorption samples (e.g., lipid droplets, protein aggregates), whereas diffused molecules show weaker signals. Secondly, the current transmission implementation constrains illumination NA by free space geometry. A reflection-mode PRIDT using water- or oil-immersion objectives could increase NA and improve spatial resolutions. Thirdly, extending the pump wavelengths into short-wave infrared regime could increase the penetration depths for thicker specimens, enabling label-free tissue imaging in vivo.

## 4. Materials and methods

### 4.1 PRIDT instrumentation and synchronization

The PRIDT system integrates a mid-IR tunable quantum cascade laser (QCL) as the pump source, a custom-built 16-angle visible laser diode array as the probe illumination module, and a high-speed camera for synchronized image acquisition. The mid-IR pump beam was generated by a nanosecond QCL (Daylight Solutions, MIRcat), operating in the mid-IR regime. The output was directed through an off-axis parabolic mirror (Thorlabs, MPD019-P01, reflected focal length 25.4 mm) to loosely focus the beam onto the sample in a transmission geometry. The resulting mid-IR illumination area is approximately 50 × 50 µm. The QCL was operated at a repetition rate of 50 kHz with a pulse width of 500 ns, parameters optimized by thermal calibration to maximize photothermal signal contrast.

The probe illumination module is a 16-channel 450-nm laser diode array arranged in a ring-shaped 3D-printed frame, which enables evenly distributed oblique illumination from 16 azimuthal angles (NA = 0.9). Each laser diode was coupled into a multimode optical fiber (NA = 0.22, 105 µm core diameter) equipped with a fiber collimator (Thorlabs, F950FC-A) to deliver collimated probe light with sufficient power density. The transmitted light was collected by a water dipping objective (Olympus, LUMPLANFLN 60X) and a tube lens (TTL200-A, focal length 200 mm). To match the illumination NA, the back pupil plane was relayed using a 4-f system consisting of lenses with focal lengths of 180 mm and 300 mm. An adjustable aperture was placed at the relayed pupil plane to restrict the effective collection NA to 0.9. The 4-f system also provided the magnification required by spatial sampling rate determined by the camera pixel size of 12 µm.

The transmitted probe signals were recorded by a CMOS camera (Adimec Q-2HFW-CXP, full-well capacity 2 million electrons), operated at kilohertz frame rates to capture images across 16 illumination angles and two pump–probe delay states. The timing of all devices was synchronized by a custom-designed MATLAB automation program. A master clock generated by an Atmel microcontroller distributed sequential trigger pulses to the 16 probe lasers using a multiplexer. A pulse generator synchronized the pump and probe lasers with the camera, alternating between on-resonance-delay and off-resonance-delay states to encode photothermal modulation. The probe pulse width was set to 500 ns to match the thermal excitation and relaxation time scale. Each camera frame accumulated 100 probe pulses to accommodate the camera well depth, which resulted in a camera frame rate of 500 Hz and a volumetric imaging speed of 15 Hz.

### 4.2 Image reconstruction

Under the first-Born approximation, the refractive index change $\Delta n$ of the scatters induced by photothermal heating modulated the transmitted probe light intensity [57]. For an oblique illumination at angle $\theta$, the transmitted intensity modulation captured by the camera is given by convolution of the 3D IDT transfer functions $H_\theta$ and $\Delta n$.

$$\widetilde{\delta I}_\theta(k_x, k_y) = H_{\theta(k_x,k_y,k_z)} \cdot \widetilde{\Delta n}(k_x, k_y, k_z) \tag{2}$$

The IDT transfer function is

$$H_{\theta(k_x,k_y,k_z)} = P(\mathbf{k_{in}})\delta(|\mathbf{k}| - k_0) \tag{3}$$

Here, $P(\mathbf{k_{in}})$ is the pupil function under the incident wavevector of $\mathbf{k_{in}}$, and $k_0 = 2\pi n_0/\lambda$ is the wavenumber of the medium.

With the measurements of 16 illumination angles at two delay states ($I_\theta^{(t)}(k_x, k_y)$), we performed IDT reconstruction at each delay state. The 2D measurements were mapped into 3D

$k$-space based on $k_z = \sqrt{k_0^2 - (k_x + k_{x,\theta})^2 - (k_y + k_{y,\theta})^2}$. The 3D refractive index was then reconstructed using Tikhonov regularization [41, 58].

$$\widetilde{n^{(t)}}(k_x, k_y, k_z) = \frac{\sum_\theta H_\theta^*(k_x, k_y, k_z) \cdot \widetilde{I_\theta^{(t)}}(k_x, k_y)}{\sum_\theta |H_\theta(k_x, k_y, k_z)|^2 + \lambda} \quad (4)$$

Here, $\lambda$ is the Tikhonov regularization term to suppress noise. Finally, the PRIDT volume was obtained by subtracting the reconstructions at two delay states.

$$\Delta n_{PRIDT}(x, y, z) = n^{(t_1)}(x, y, z) - n^{(t_2)}(x, y, z) \quad (5)$$

For hyperspectral imaging, this process was repeated at each QCL wavelength to generate volumetric chemical maps.

### 4.3 Cellular sample preparation

OVCAR5 cells were cultured in RPMI-1640 medium supplemented with 10% (v/v) fetal bovine serum (FBS) and 1% (v/v) penicillin-streptomycin (P/S) at 37 °C in a humidified incubator containing 5% $CO_2$. For live-cell imaging, cells were seeded onto $CaF_2$ slides and incubated overnight, followed by three washes with phosphate-buffered saline (PBS). A live-cell imaging buffer was used to maintain cellular activity during imaging. For azide-labeled samples, cells were first seeded on $CaF_2$ slides and incubated overnight in standard culture medium, then treated with azido-palmitic acid (final concentration: 50 μM) in RPMI-1640 medium containing 1% (v/v) FBS and 1% (v/v) P/S for 24 h. After treatment, cells were washed three times with PBS and fixed with 4% formalin.

### 4.4 Temperature measurements

Temperature measurements were performed by a custom-built pump-probe imaging system incorporating thermo-sensitive fluorescence modulation as the probe. The setup used a counter-propagating geometry, with the pump laser pulses operating at the same repetition rate (50 kHz) and pulse width (500 ns) as in the PRIDT imaging configuration. The probe system was implemented by a wide-field fluorescence microscope with a 488-nm laser (Cobolt, 06-MLD), modulated to a pulse width of 500 ns as the excitation source, and the reflected probe fluorescence was recorded by a sCMOS camera (Andor, ZYLA-5.5-USB3- S). FITC dye was selected as the thermo-sensitive probe.

To calibrate the fluorescence thermo-sensitivity, FITC–water solutions were placed on a temperature- controlled plate. Fluorescence intensity was recorded while tuning the temperature, thus providing a calibration curve of the temperature-dependent fluorescence response. For in situ measurements of background heating induced by accumulated QCL excitation, the pump and probe were synchronized so that pump pulse trains alternated between "Hot" and "Cold" states. In the Hot states, the pump-probe delay was tuned to 18 μs to capture the residual thermal relaxation. The photothermal signals were then extracted by subtracting the Hot and Cold frames. Each camera frame integrated 1000 probe pulses, and 5 frames were averaged to improve the SNR. The fluorescence modulation was calibrated to account for photobleaching effects. Finally, the measured photothermal signals were converted into temperature rise using the pre-calibrated FITC thermo-sensitivity curve.

### 4.5 Hyperspectral data acquisition and processing

Hyperspectral data were collected by tuning the QCL wavelength to targeted molecular resonances. To synchronize wavelength tuning with multi-angle image acquisition, TTL signals were used as wavelength triggers to initiate each imaging cycle at a given wavenumber. At each QCL wavelength, 5 cycles of 32 camera frames were recorded and averaged to reconstruct the corresponding PRIDT volume. The wavelength scan introduced a dead time of approximately 200 ms between adjacent wavenumbers. For applications requiring faster spectral acquisition, a continuous wavelength sweep mode without dead time can be used. During long image acquisition processes, there were motion-induced artifacts in the surrounding medium. To mitigate these artifacts in the hyperspectral image stacks and attain quantitative chemical abundance maps, we applied a least-square unmixing method.

$$\mathbf{I}(x,y,z) = \sum_i a_i(x,y,z)\mathbf{e_i} + \varepsilon \qquad (6)$$

Here, $\mathbf{I}_{x,y,z}$ is the pixel-wise spectrum, $\mathbf{e}_i$ represents the normalized spectral basis vectors within the wavelengths of interest, $a_i$ is the quantitative chemical abundance map to be retrieved, and $\varepsilon$ accounts for residual signals from noise or other chemical contents. Then, abundance maps were recovered by solving a least-square problem.

$$a_i(x,y,z) = \arg\min_a ||\mathbf{I}(x,y,z) - \mathbf{E} \cdot a||^2 \qquad (7)$$

Here, $\mathbf{E}$ is a matrix that contains all normalized spectral basis vectors of interest. The hyperspectral depth-resolved image stacks were reconstructed as the calculated abundance distributions.

### *4.6 Video data acquisition and processing*

PRIDT video of living cells was recorded at single mid-IR excitation wavelength (1744 cm$^{-1}$) to match the C=O vibrations of lipid droplets. To achieve video-rate volumetric chemical imaging, each chemical volume was reconstructed directly from 32 camera frames without frame averaging. To suppress noise in the time-resolved volumetric data $I$ (x, y, z, t), we applied the block-matching 4D (BM4D) filtering algorithm. For long time-lapse imaging, cells were recorded every 1 min for 1 h, with each acquisition consisting of 10 cycles (32 frames × 2 delay states).

**Funding.** This work is supported by Grant No. 2023-321163 from the Chan Zuckerberg Initiative Donor-Advised Fund at the Silicon Valley Community Foundation.

**Acknowledgment.** The authors thank Dr. Qing Xia from Electrical and Computer Engineering Department at Boston University for providing the thermal sensitivity calibration of fluorescence, and Guo Chen from Electrical and Computer Engineering Department at Boston University for help with 3D printing of the probe laser holder.

Author Contributions. J.-X.C. and L.T. conceived and supervised the project. D.J. and D.D. designed and built the PRIDT system. D.J. and T.L. developed and performed the image processing pipeline. H.Z. designed the illumination module. J.Z. contributed to the IDT reconstruction algorithm. X.T. provided the biological samples. D.J. wrote the manuscript with input from all authors.

**Disclosures.** The authors declare no conflicts of interest.

**Data availability.** All data are available in the main text or the supplementary documents.

**Supplemental documents.** See Supplement 1 for supporting content.

## References

1. Lichtman, J. W. & Conchello, J.-A. Fluorescence microscopy. *Nat. Methods* **2**, 910–919 (2005).
2. Webb, R. H. Confocal optical microscopy. *Reports on Prog. Phys.* **59**, 427 (1996).
3. Denk, W., Strickler, J. H. & Webb, W. W. Two-Photon Laser Scanning Fluorescence Microscopy. *Science* **248**, 73–76 (1990).
4. Helmchen, F. & Denk, W. Deep tissue two-photon microscopy. *Nat. Methods* **2**, 932–940 (2005).
5. Huisken, J., Swoger, J., Del Bene, F., Wittbrodt, J. & Stelzer, E. H. K. Optical Sectioning Deep Inside Live Embryos by Selective Plane Illumination Microscopy. *Science* **305**, 1007–1009 (2004).
6. Keller, P. J., Schmidt, A. D., Wittbrodt, J. & Stelzer, E. H. Reconstruction of Zebrafish Early Embryonic Development by Scanned Light Sheet Microscopy. *Science* **322**, 1065–1069 (2008).
7. Tokunaga, M., Imamoto, N. & Sakata-Sogawa, K. Highly inclined thin illumination enables clear single-molecule imaging in cells. *Nat. Methods* **5**, 159–161 (2008).
8. Levoy, M., Ng, R., Adams, A., Footer, M. & Horowitz, M. Light field microscopy. *ACM Trans. Graph.* **25**, 924–934 (2006).
9. Hua, X., Liu, W. & Jia, S. High-resolution Fourier light-field microscopy for volumetric multi-color live-cell imaging. *Optica* **8**, 614–620 (2021).
10. Cheng, J.-X. & Xie, X. S. Vibrational spectroscopic imaging of living systems: An emerging platform for biology and medicine. *Science* **350**, aaa8870 (2015).
11. Cheng, J.-X. & Xie, X. S. Coherent Anti-Stokes Raman Scattering Microscopy: Instrumentation, Theory, and Applications. *The J. Phys. Chem. B* **108**, 827–840 (2004).
12. Freudiger, C. W. *et al.* Label-Free Biomedical Imaging with High Sensitivity by Stimulated Raman Scattering Microscopy. *Science* **322**, 1857–1861 (2008).
13. Hu, F., Shi, L. & Min, W. Biological imaging of chemical bonds by stimulated Raman scattering microscopy. *Nat. Methods* **16**, 830–842 (2019).
14. Chen, X. *et al.* Volumetric chemical imaging by stimulated Raman projection microscopy and tomography. *Nat. Commun.* **8**, 15117 (2017).
15. Lin, P. *et al.* Volumetric chemical imaging in vivo by a remote-focusing stimulated Raman scattering microscope. *Opt. Express* **28**, 30210–30221 (2020).
16. Qian, C. *et al.* Super-resolution label-free volumetric vibrational imaging. *Nat. Commun.* **12**, 3648 (2021).
17. Gao, X., Li, X. & Min, W. Absolute Stimulated Raman Cross Sections of Molecules. *The J. Phys. Chem. Lett.* **14**, 5701–5708 (2023).
18. Zhang, D. *et al.* Depth-resolved mid-infrared photothermal imaging of living cells and organisms with submicrometer spatial resolution. *Sci. Adv.* **2**, e1600521 (2016).
19. Li, Z., Aleshire, K., Kuno, M. & Hartland, G. V. Super-Resolution Far-Field Infrared Imaging by Photothermal Heterodyne Imaging. *The J. Phys. Chem. B* **121**, 8838–8846 (2017).
20. Zhang, D. *et al.* Bond-selective transient phase imaging via sensing of the infrared photothermal effect. *Light. Sci. & Appl.* **8**, 116 (2019).
21. Samolis, P. D. & Sander, M. Y. Phase-sensitive lock-in detection for high-contrast mid-infrared photothermal imaging with sub-diffraction limited resolution. *Opt. Express* **27**, 2643–2655 (2019).
22. Yin, J. *et al.* Video-rate mid-infrared photothermal imaging by single-pulse photothermal detection per pixel. *Sci. Adv.* **9**, eadg8814 (2023).
23. Bai, Y., Yin, J. & Cheng, J.-X. Bond-selective imaging by optically sensing the mid-infrared photothermal effect. *Sci. Adv.* **7**, eabg1559 (2021).
24. Xia, Q., Yin, J., Guo, Z. & Cheng, J.-X. Mid-Infrared Photothermal Microscopy: Principle, Instru- mentation, and Applications. *The J. Phys. Chem. B* **126**, 8597–8613 (2022).
25. Teng, X. *et al.* Mid-infrared Photothermal Imaging: Instrument and Life Science Applications. *Anal. Chem.* **96**, 7895–7906 (2024).
26. Furstenberg, R., Kendziora, C. A., Papantonakis, M. R., Nguyen, V., McGill, R. A. Chemical imaging using infrared photothermal microspectroscopy. *Proc. SPIE.* **8374**, 837411 (2012).
27. Aleshire, K., *et al.* Far-field midinfrared superresolution imaging and spectroscopy of single high aspect ratio gold nanowires. *Proc. Natl. Acad. Sci.* **117**, 2288-2293 (2020).
28. Bai, Y. *et al.* Ultrafast chemical imaging by widefield photothermal sensing of infrared absorption. *Sci. Adv.* **5**, eaav7127 (2019).
29. Zong, H. *et al.* Background-Suppressed High-Throughput Mid-Infrared Photothermal Microscopy via Pupil Engineering. *ACS Photonics* **8**, 3323–3336 (2021).
30. Schnell, M. *et al.* All-digital histopathology by infrared-optical hybrid microscopy. *Proc. Natl. Acad. Sci.* **117**, 3388–3396 (2020).
31. Fu, P. *et al.* Breaking the diffraction limit in molecular imaging by structured illumination mid-infrared photothermal microscopy. *Adv. Photonics* **7**, 036003 (2025).
32. Lee, D. *et al.* Wide-field bond-selective fluorescence imaging: from single-molecule to cellular imaging beyond video rate. *Optica* **12**, 148–157 (2025).
33. Zhang, Y. *et al.* Fluorescence-Detected Mid-Infrared Photothermal Microscopy. *J. Am. Chem. Soc.* **143**, 11490–11499 (2021).

34. Li, M. et al. Fluorescence-Detected Mid-Infrared Photothermal Microscopy. *J. Am. Chem. Soc.* **143**, 10809–10815 (2021).
35. Jia, D. *et al.* 3D Chemical Imaging by Fluorescence-detected Mid-Infrared Photothermal Fourier Light Field Microscopy. *Chem. & Biomed. Imaging* **1**, 260–267 (2023).
36. Popescu, G. *Quantitative Phase Imaging of Cells and Tissues* (McGraw-Hill Education, 2011), 1st edition edn.
37. Choi, W. *et al.* Tomographic phase microscopy. *Nat. Methods* **4**, 717–719 (2007).
38. Sung, Y. *et al.* Optical diffraction tomography for high resolution live cell imaging. *Opt. Express* **17**, 266–277 (2009).
39. Cotte, Y. *et al.* Marker-free phase nanoscopy. *Nat. Photonics* **7**, 113–117 (2013).
40. Dong, D. *et al.* Super-resolution fluorescence-assisted diffraction computational tomography reveals the three-dimensional landscape of the cellular organelle interactome. *Light. Sci. & Appl.* **9**, 11 (2020).
41. Ling, R., Tahir, W., Lin, H.-Y., Lee, H. & Tian, L. High-throughput intensity diffraction tomography with a computational microscope. *Biomed. Opt. Express* **9**, 2130–2141 (2018).
42. Li, J. *et al.* High-speed in vitro intensity diffraction tomography. *Adv. Photonics* **1**, 066004 (2019).
43. Li, J. *et al.* Transport of intensity diffraction tomography with non-interferometric synthetic aper- ture for three-dimensional label-free microscopy. *Light. Sci. & Appl.* **11**, 154 (2022).
44. Tamamitsu, M., Toda, K., Horisaki, R. & Ideguchi, T. Quantitative phase imaging with molecular vibrational sensitivity. *Opt. Lett.* **44**, 3729–3732 (2019).
45. Tamamitsu, M. *et al.* Label-free biochemical quantitative phase imaging with mid-infrared photother- mal effect. *Optica* **7**, 359–366 (2020).
46. Tamamitsu, M. *et al.* Mid-infrared wide-field nanoscopy. *Nat. Photonics* **18**, 738–743 (2024).
47. Zhao, J. *et al.* Bond-selective intensity diffraction tomography. *Nat. Commun.* **13**, 7767 (2022).
48. Yin, J. *et al.* Nanosecond-resolution photothermal dynamic imaging via MHz digitization and match filtering. *Nat. Commun.* **12**, 7097 (2021).
49. Samolis, P. D., Zhu, X. & Sander, M. Y. Time-Resolved Mid-Infrared Photothermal Microscopy for Imaging Water-Embedded Axon Bundles. *Anal. Chem.* **95**, 16514–16521 (2023).
50. Bolarinho, R., Yin, J., Ni, H., Xia, Q. & Cheng, J.-X. Background-Free Mid-Infrared Photothermal Microscopy via Single-Shot Measurement of Thermal Decay. *Anal. Chem.* **97**, 4299–4307 (2025).
51. Sletten, E. M. & Bertozzi, C. R. Bioorthogonal Chemistry: Fishing for Selectivity in a Sea of Functionality. *Angewandte Chemie Int. Ed.* **48**, 6974–6998 (2009).
52. Hang, H. C. *et al.* Chemical Probes for the Rapid Detection of Fatty-Acylated Proteins in Mammalian Cells. *J. Am. Chem. Soc.* **129**, 2744–2745 (2007).
53. Bai, Y. *et al*. Single-cell mapping of lipid metabolites using an infrared probe in human-derived model systems. *Nat. Commun.* **15**, 350 (2024).
54. De Clercq, E. HIV resistance to reverse transcriptase inhibitors. *Biochem. Pharmacol.* **47**, 155–169 (1994).
55. De Clercq, E. & Li, G. Approved Antiviral Drugs over the Past 50 Years. *Clin. Microbiol. Rev.* **29**, 695–747 (2016).
56. Wang, X.-d., Wolfbeis, O. S. & Meier, R. J. Luminescent probes and sensors for temperature. *Chem. Soc. Rev.* **42**, 7834–7869 (2013).
57. Wolf, E. Three-dimensional structure determination of semi-transparent objects from holographic data. *Opt. Commun.* **1**, 153–156 (1969).
58. Tian, L. & Waller, L. 3D intensity and phase imaging from light field measurements in an LED array microscope. *Optica* **2**, 104–111 (2015).